\documentclass[%
 reprint,
 amsmath,amssymb,
 aps,
prb,
]{revtex4-2}

\usepackage{graphicx}% Include figure files
\usepackage{dcolumn}% Align table columns on decimal point
\usepackage{bm}% bold math
\usepackage{color}
\newcommand{\BibitemShut}[1]{.}

\begin{document}

\title{NMR evidence for energy gap opening in thiol-capped platinum nanoparticles}
\author{Takuto Fujii$^1$} 
\author{Kaita Iwamoto$^1$}
\author{Yusuke Nakai$^1$} 
\email{nakai@sci.u-hyogo.ac.jp}
\author{Taisuke Shiratsu$^1$} 
\author{Hiroshi Yao$^2$}
\author{Koichi Ueda$^1$} 
\author{Takeshi Mito$^1$}
\affiliation{$^1$Graduate School of Science, University of Hyogo, 3-2-1 Kouto, Kamigori-cho, Ako-gun, Hyogo 678-1297, Japan, 
$^2$Division of Chemistry for Materials, Graduate School of Engineering, Mie University, 1577 Kurimamachiya-cho, Tsu, Mie 514-8507, Japan} 

\date{\today}% It is always \today, today,
             %  but any date may be explicitly specified

\begin{abstract}
When the particle size of metal is reduced, it is expected that an energy gap will open due to the quantum size effect. However, the energy gap in platinum (Pt) metal nanoparticles has not been observed directly by nuclear magnetic resonance (NMR). To investigate particle size dependence of the electronic state of Pt nanoparticles, we performed $^{195}$Pt-NMR experiments on thiol-capped Pt nanoparticles with three different average diameters of less than 3 nm. For the nanoparticles with a diameter of 2.8 nm, we observed usual metallic behavior with a smaller density of states than that of the bulk Pt. In contrast, the temperature dependence of $1/T_1T$ in the nanoparticles less than 2.5 nm in diameter is an activation-energy form above 150 K, which is semiconducting behavior with an energy gap of the order of 2000 K.
 The significant decrease in $1/T_{\rm 1}T$ by more than two orders of magnitude in the smaller Pt nanoparticles compared to the bulk Pt is attributable to the disappearance of the density of states at the Fermi energy, which is consistent with the opening of an energy gap. These results indicate the metal-insulator transition below 2.5 nm in diameter is present in our thiol-capped Pt nanoparticle samples. The effect of the thiol-capping on the electronic structure suggested by the experimental results is also discussed.
\end{abstract}

\maketitle

At the nanometer scale in solids, the change in electronic states due to the reduction in size leads to fascinating physical properties that are completely different from those of bulk materials \cite{Alivisatos1996}. In metallic nanoparticles, the quantum size effect, also known as the Kubo effect, predicts that as the size decreases, the energy levels of the electrons become discrete due to spatial confinement, and the energy gap opens at the Fermi level \cite{Kubo1962,KuboReview1984}. In addition to the quantum size effects, surface effects are also a very interesting subject, and various intriguing physical properties emerge in nano-sized systems depending on capping molecules (ligands) \cite{Crespo2004,Garcia2007}. A clear understanding of the quantum size effect and the surface effect could open the way for the precise control of nanoscale properties, leading to a wider range of material applications.

Nuclear magnetic resonance (NMR) is a useful microscopic tool to probe the electronic structure in solids and has been applied to characterize the properties of metallic nanoparticles \cite{Slicter1986,Halperin1986,Goto1989,Klink2000,Marbella2015}. Among the metallic nanoparticles,  platinum ($^{195}$Pt) is the most studied nucleus because of favorable NMR properties such as relatively high natural abundance (33.8\%), its nuclear spin $1/2$ that excludes quadrupolar effects, and a moderately high gyromagnetic ratio.  Furthermore, Pt nanoparticles arguably represent a very important class of materials because they play important roles in many catalytic processes \cite{Seh2017}. Understanding their electronic properties is essential for maximizing the catalytic properties of the Pt nanoparticles.  

It has been experimentally established that the $^{195}$Pt NMR spectra of the Pt nanoparticles are very broad, depending on the particle size; the $^{195}$Pt NMR peak corresponding to the bulk Pt metal diminishes in intensity as the particle size decreases, and a surface peak appears around $K\sim0\%$. These spectral features are commonly seen in many kinds of the Pt nanoparticles in the literature, including oxide-supported Pt nanoparticles, Pt cluster compounds surrounded by ligands, and Pt nanoparticles in zeolite matrices \cite{Slichter1981,Rhodes1982i,Rhodes1982ii,Makowka1982,Klink1984,Makowka1985,Tong1993,Tong1995,Putten1993,Fritschij1999,Yu1980,Yu1993}.

Despite many NMR studies of the Pt nanoparticles, the activated temperature dependence of the nuclear spin-lattice relaxation rate $1/T_1$ expected from the appearance of an energy gap has not been observed for the Pt nanoparticles. In Pt$_{309}$Phen$^{\ast}_{36}$O$_{30}$,  although the authors of Refs.\onlinecite{Putten1993} and \onlinecite{ Fritschij1999} did not observe gapped behavior in the temperature dependence of $1/T_1$, they estimated an energy gap indirectly by fitting a $T_1$ relaxation curve to a theoretical model based on random matrix approximation. However, there remains an unresolved inconsistency that their model cannot explain the temperature-independent $^{195}$Pt NMR spectrum. Recently, a $^{195}$Pt NMR study on Pt$_{13}$ nanoclusters with average-size of 0.73 nm covered with trialkyl aluminum ligands showed that $T_1$ is four orders magnitude longer than $T_1$ for the bulk Pt \cite{Rees2013}, indicating nonmetallic nature of the Pt$_{13}$ nanoclusters, although no temperature dependence of $1/T_1$ was reported. Quite recently, the anomalous size-dependent magnetic fluctuations found in the Pt nanoparticles coated with polyvinylpyrrolidone (PVP) with an average diameter larger than 2.5 nm have been interpreted in relation to the discrete energy levels expected from the quantum size effect \cite{Okuno2020}. However, no measurements were made in the region below 2.5 nm, where the quantum size effect should be more pronounced.

In this paper, we examine the quantum size effects by using $^{195}$Pt-NMR measurements on the well-characterized, thiol-capped monodisperse Pt nanoparticles with the particle sizes ranging from 2.8 nm to 2.1 nm in diameter. The quantum size effects can significantly increase $T_1$ at low temperatures due to a gap, which has hindered systematic NMR measurements on small metal nanoparticles so far. Here, a modern NMR spectrometer enabled us to measure an extremely long $T_1$ in the Pt nanoparticles with an average particle diameter below 3 nm down to 20 K. 

Three samples of the Pt nanoparticles with average diameters of 2.1 $\pm$0.5 nm, 2.5$ \pm$0.7 nm, and 2.8$ \pm$0.6 nm, hereafter denoted as Pt(2.1), Pt(2.5), and Pt(2.8) respectively, were measured. The samples were prepared by chemically reducing the solution containing H$_{2}$PtCl$_{6}$ and Mercaptosuccinic acid (MSA) \cite{ChenKimura2001}. The particle size, controlled by the ratio of H$_{2}$PtCl$_{6}$ and MSA, was determined by using TEM measurement (see Supplemental Material). $^{195}$Pt NMR experiments were carried out with a phase-coherent pulsed spectrometer (Thamway Co., Ltd.) in 6.5 T and 9.4 T  superconducting magnets. The $^{195}$Pt NMR spectra were obtained by frequency-sweep with a spin-echo pulse sequence in the static magnetic field. The obtained NMR spectra were corrected for the effect of spin-spin relaxation time $T_2$ by measuring $T_{2}$ over the spectra (see Fig.S4 in Supplemental Material). $T_1$ was measured with a saturation method. The typical nuclear magnetization curves are shown in Fig. 2. We measured the temperature dependence of $T_1$ at the peak position with the Knight shift $K\sim 0.2 \%$,  which is almost the same position as that of Ref. \onlinecite{Fritschij1999}. 

%----------------------------Fig.1----------------------------------------
\begin{figure}[tbp]
\centering
\includegraphics[clip,scale=0.33]{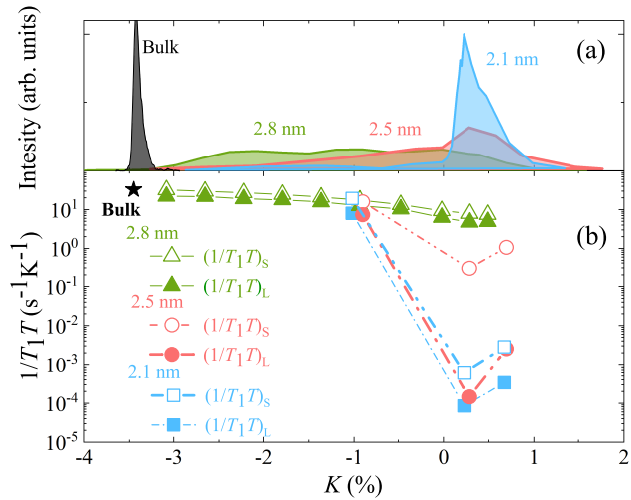}
\caption{$K$ dependence of $1/T_1T$ for Pt(2.1), Pt(2.5), Pt(2.8), and the bulk Pt metal at 40 K (a), along with their $^{195}$Pt NMR spectra (b).}
\label{Fig2a}
\end{figure}
%----------------------------Fig.1----------------------------------------
Figure 1(a) shows the $^{195}$Pt NMR spectra for Pt(2.8), Pt(2.5), and Pt(2.1) at 40 K. As a reference, Fig. 1 also shows $^{195}$Pt NMR spectrum of the bulk Pt (99.9\% purity and 200 mesh, Kishida Chemical Co., Ltd.) at 40 K. Similar to previous studies on the Pt nanoparticles \cite{Slichter1981,Rhodes1982i,Rhodes1982ii,Makowka1982,Klink1984,Makowka1985,Tong1993,Tong1995,Putten1993,Fritschij1999}, the spectra are broad and the line shape is strongly dependent on  the average particle size. As the particle size is reduced, the intensity near the bulk Pt position ($\sim-3.5\%$) gets weaker, while the intensity around $K=0\%$ attributed to the surface Pt atoms gets greater \cite{Putten1993,Fritschij1999,Tong2005}. This indicates that the volume fraction of the thiol-capped nanoparticles with an electronic state similar to that of the bulk Pt is negligible. The spectrum of Pt(2.5) is similar to that observed in the metal cluster compound Pt$_{309}$Phen$^{\ast}_{36}$O$_{30}$ \cite{Putten1993,Fritschij1999}, and the more symmetric spectrum of Pt(2.1) is similar to that of the Pt$_{13}$ nanoclusters covered with trialkyl aluminum \cite{Rees2013}. Note that our thiol-capped nanoparticles do not show the spectrum characteristics of Pt oxides (PtO$_2$ near $K \sim 1 \%$) found in the oxide-supported Pt nanoparticles \cite{Bucher1989}, most probably owing to the thiol-capping that avoids surface oxidation.

%----------------------------Fig.2----------------------------------------
\begin{figure*}[ht]
\centering
\includegraphics[scale=0.35]{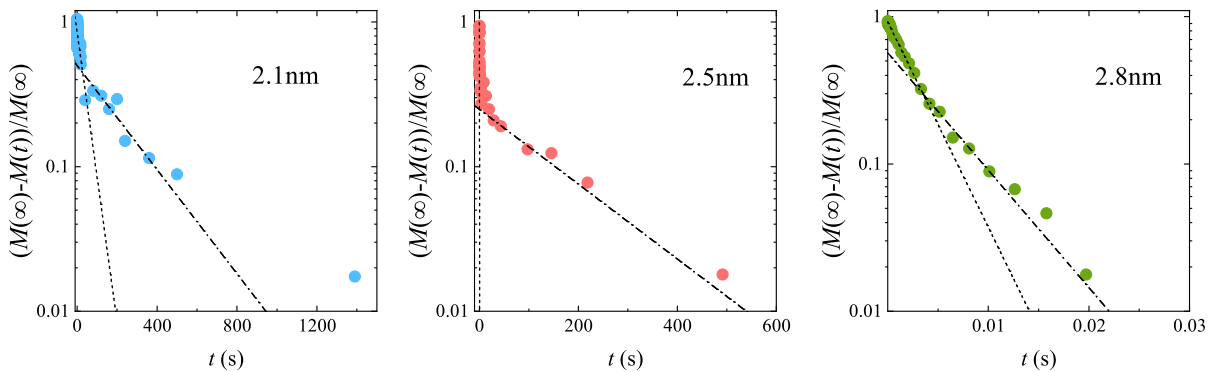}
\caption{$T_1$ relaxation curves for Pt(2.1), Pt(2.5), and Pt(2.8) at $K=0.2 \%$ and 40 K. The dotted and dashed-and-dotted lines represent the short and long components of $T_1$, respectively (see text).}
\label{Fig1}
\end{figure*}
%----------------------------Fig.2----------------------------------------
To examine the density of states at the Fermi energy $E_F$, $D(E_F)$, that can reflect the existence of an energy gap, we measured the temperature dependence of $1/T_1$. For nonmagnetic materials, $1/T_1$ is a good measure of $D(E_F)^2$. In the bulk Pt, the $T_1$ relaxation curve follows a single exponential because $^{195}$Pt nuclear spin has a nuclear magnetic moment of $I$ = 1/2. In the case of nanoparticles, it is expected that the relaxation curve can be non-single exponential because the relaxation time is determined by the ratio of local densities of state of the $s$-electron and the $d$-electron contributions and/or the randomness of the particle size in the sample. However, only a few reports have observed such behavior possibly due to an ultralong $T_1$ in small nanoparticles \cite{Klink2000}. In this study, we observed obvious two-component behavior with a longer $T_1$ component than 100 s below 100 K in Pt(2.1). Figure 2 shows the $T_1$ relaxation curves for Pt(2.8), Pt(2.5), and Pt(2.1) at $K = 0.2 \%$ and 40 K, and they exhibit non-single exponential at all temperatures. The recovery curves of the nuclear magnetization $M(t)$ with $t$ after a saturation pulse are then fitted successfully with two components of $T_1$, a short $T_1$ component, $T_{\rm 1S}$, and a long $T_1$ component, $T_{\rm 1L}$, as follows: 
\begin{equation}
\label{eq:1}
\frac{M(\infty)-M(t)}{M(\infty)}=C\exp\left(\frac{-t}{T_{\rm 1S}}\right)+(1-C)\exp\left(\frac{-t}{T_{\rm 1L}}\right).
\end{equation}
Here, the fitting parameter $C$ was treated as temperature independent with a value of 0.58 for Pt(2.8), 0.5 for Pt(2.5), and 0.54 for Pt(2.1). 

Figure 1(b) shows the $K$ dependence of $1/T_{\rm 1}T$, which represents the variations in $D(E_F)$ with respect to $K$. In Pt(2.8), the value of $1/T_{\rm 1}T$ around $K=-3\%$ is close to that of the bulk Pt, while $1/T_{\rm 1}T$ decreases with increasing $K$ up to around $K=0\%$. For Pt(2.5) and Pt(2.1), $1/T_{\rm 1}T$ around $K= 0\%$ shows a value more than four orders of magnitude smaller than that of Pt(2.8), which indicates the significant reduction in the local $D(E_F)$ below the particle diameter of 2.5 nm. 
%----------------------------Fig.3----------------------------------------
\begin{figure}[tbp]
\centering
\includegraphics[clip,scale=0.36]{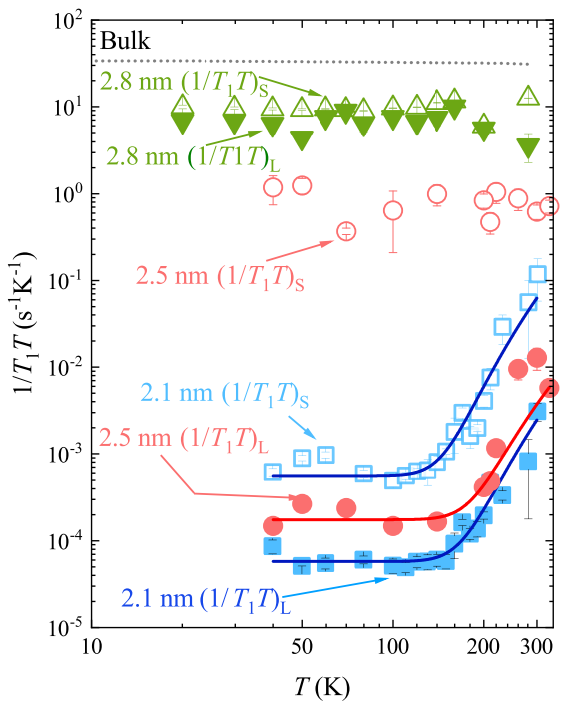}
\caption{Temperature dependence of $(T_1T)^{-1}$ for Pt(2.1), Pt(2.5), and Pt(2.8). The dotted line represents the temperature dependence of $(T_1T)^{-1}$ for the bulk Pt metal, which is cited from Ref. \onlinecite{MetallicShifts}. The solid lines are fits for the experimental data of the nanoparticles (see texts in detail). }
\label{Fig3}
\end{figure}
%----------------------------Fig.3----------------------------------------

To shed light on the origin of the sudden decrease in $1/T_1T$, the temperature dependence of $1/T_1T$  at $K\sim0.2\%$ is shown in Fig. 3. In Pt(2.8), $(1/T_{\rm 1L}T)$ and $(1/T_{\rm 1S}T)$  are one-third of the bulk value and nearly temperature-independent. Note that for the bulk Pt metal, the temperature dependence of $1/T_1T$ is constant below 300 K \cite{MetallicShifts,Shaham1978}, which is the typical behavior of the metal. This result indicates the metallic nature of  Pt(2.8) with a reduced $D(E_F)$, which is consistent with the previous measurements on similar particle sizes \cite{Bucher1988}.
In fact, by calculating the $D(E_F)$ derived from $1/T_1T$ and $K$ according to the method considering many-body effects in Ref. \cite{Bucher1988}, we estimated that the $D(E_F)$ of $s$- and $d$-electrons is reduced by about 70\% and 35\%, respectively, compared to the bulk.

In contrast, there is a large difference between $(1/T_{\rm 1S}T)$ and $(1/T_{\rm 1L}T)$ in Pt(2.5); the longer component $(1/T_{\rm 1L}T)$ is approximately $1000$ times smaller than $(1/T_{\rm 1S}T)$ at low temperatures. Furthermore, $(1/T_{\rm 1S}T)$ in Pt(2.5) is nearly two orders of magnitude smaller than that of the bulk Pt, suggesting that $D(E_F)$ in Pt(2.5) is much smaller than the bulk Pt. 

Notably, $1/T_1T$ of Pt(2.1) and $(1/T_{\rm 1L}T)$ of Pt(2.5) increases rapidly on heating, exhibiting semiconductor-like behavior above 150K. We attribute the increase in $(1/T_1T)$ to an increase in the number of carriers thermally excited across a gap induced by the quantum size effect. In contrast, we observe weak metallic behavior  in these samples at low temperatures. This result suggests that the dominant contributions to $T_1$ from the energy bands differ depending on the temperature range. Namely, $T_1$ relaxation is induced by an extremely low $D(E_F)$ at low temperature,  which originates from a band faintly overlapping the Fermi level,  while the thermally activated carriers across the energy gap with a larger $D(E)$ determine $T_1$ at high temperature. In this case, $1/T_1T$ is given by \cite{Bloembergen1954,Selbach1979} 
\begin{equation}
\label{eq:2}
\frac{1}{T_{1}T}=AT\exp\left(\frac{-\Delta_{\rm NMR}}{2k_BT}\right)+\left(\frac{1}{T_{1}T}\right)_{\rm const}, 
\end{equation}
where $A$ is a constant and $\left(\frac{1}{T_{1}T}\right)_{\rm const}$ is the remaining metallic contribution at low temperatures. The fitting results plotted by the solid lines in Fig. 3 closely reproduce the experimental results. The estimated energy gap $\Delta_{\rm NMR}$ is shown in Fig. 4(a). Figure 4(a) shows, when the diameter of the thiol-protected Pt nanoparticles is less than 2.5 nm, a metal-insulator transition occurs and an energy gap of the order of 2000 K appears.
%----------------------------Fig.4----------------------------------------
\begin{figure}[tbp]
\centering
\includegraphics[clip,scale=0.32]{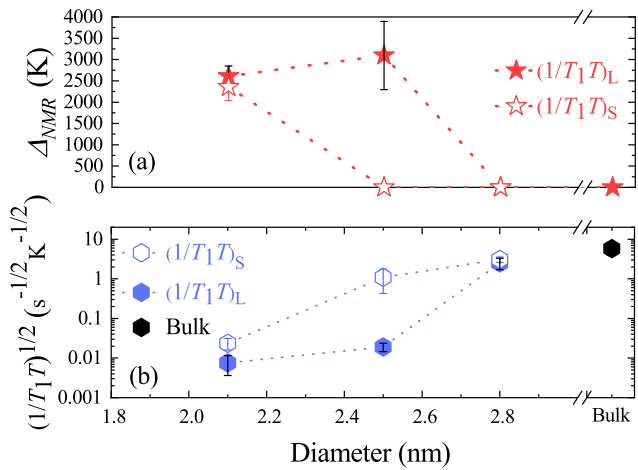}
\caption{Average diameter dependence of (a) the energy gap determined by NMR, $\Delta$$_{\rm NMR}$, and (b) $1/T_1T$ at the lowest temperature with lines as a guide to the eye.}
\label{Fig4}
\end{figure}
%----------------------------Fig.4----------------------------------------

In the framework of the free-electron model \cite{Halperin1986}, the averaged level spacing $\delta$
 of a given metal nanoparticle may be estimated as  
\begin{equation}
\label{eq:3}
\delta \left(\equiv \frac{1}{D(E_F)}\right) =\frac{4E_F}{3N} . 
\end{equation}
Here $N$ is the number of valence electrons in a nanoparticle. Assuming $E_F =$ 8.79 eV expected for the bulk Pt \cite{Sigalas1992} and one electron per Pt atom, $\delta$ was estimated as 251 K (423 K) for a nanoparticle with a diameter of 2.5nm (2.1nm). In our experiment, we found that $\Delta$$_{\rm NMR}$ = 3160$\pm$ 800 K for Pt(2.5)  and  2140 $\pm$ 240 K for Pt(2.1) are  more than four times as large as $\delta$. Previous density functional theory (DFT) calculations with a plane wave basis set give an energy gap of  approximately 1500 K for a bare 55-atom Pt cluster with a diameter of 1.2 nm \cite{Xiao2004}, which is the same order of magnitude as those observed in the present samples. The discrepancy between $\Delta_{\rm NMR}$ and $\delta$ is likely to be ascribed to the rough estimation using Eq.(\ref{eq:3}) and/or to the Pt-thiol bonding as described below. 

The metal-insulator transition below 2.5 nm in diameter is also indicated from the particle-size dependence of $(1/T_{\rm 1}T)^{1/2}$ at the lowest temperatures as shown in Fig. 4(b). The significant decrease in $1/T_{\rm 1}T$ by more than two orders of magnitude in Pt(2.1) and Pt(2.5) compared to the bulk Pt is attributable to the disappearance of the $D(E_F)$, indicating a threshold diameter for the appearance of an energy gap is around 2.5 nm for the present thiol-capped Pt nanoparticles. We emphasize that the measurements using the smaller samples than the threshold enable us to observe the clear gap opening and may lead the different size effects than those reported in Ref.25 where larger diameter samples were investigated \cite{Okuno2020}.

One characteristic found in Pt(2.5) and Pt(2.1) is the tiny $D(E_F)$ below 120 K, as seen in Figs. 3 and 4(b). The weak-metallic states may originate from an unavoidable particle-size distribution, which needs further investigation. Besides, further studies are needed to determine the role of surrounding capping molecules in the appearance and magnitude of the energy gap. 
Although the effect of thiol-capping on platinum nanoparticles is not fully understood, theoretically thiol-capping may induce a decrease in the $D(E_F)$ as reported in Pd nanoparticles \cite{Fresch}, a metal with similar properties to Pt. This decrease in the $D(E_F)$ can increase $\delta$ as expected from Eq.(\ref{eq:3}), which may explain the large $\Delta_{\rm NMR}$ obtained in the present experiment. In this connection, the difference in the capping molecule can be another explanation for the difference in the energy gaps estimated in Ref.25 and this study. 

In summary, we report systematic $^{195}$Pt-NMR for three different diameter samples of the thiol-capped Pt nanoparticles with an average diameter of less than 3 nm. For Pt(2.8), we observed usual metallic behavior with a smaller density of states than that of the bulk Pt. In contrast, the temperature dependence of $1/T_1T$ in Pt(2.1) and Pt(2.5) is an activation-energy form above 150 K, which is typical in a semiconductor with an energy gap of the order of 2000 K.
 The significant decrease in $1/T_{\rm 1}T$ by more than two orders of magnitude in Pt(2.1) and Pt(2.5) compared to the bulk Pt is attributable to the disappearance of the $D(E_F)$, which is consistent with the opening of an energy gap. These results indicate the metal-insulator transition below 2.5 nm in diameter is present in our thiol-capped Pt nanoparticle sample. Our NMR results are consistent with the prediction of discrete electron-energy levels from the quantum size effect. Furthermore, through the change in the density of states at the Fermi level, it is suggested that thiol-capping may be involved in determining the size of the energy gap.

\begin{acknowledgments}
We   thank   Prof. A. Miyazawa and Dr. Y. Nishino in the University of Hyogo for their helpful assistance on  the  TEM  measurements.
\end{acknowledgments}


\begin{thebibliography}{32}%
\makeatletter
\providecommand \@ifxundefined [1]{%
 \@ifx{#1\undefined}
}%
\providecommand \@ifnum [1]{%
 \ifnum #1\expandafter \@firstoftwo
 \else \expandafter \@secondoftwo
 \fi
}%
\providecommand \@ifx [1]{%
 \ifx #1\expandafter \@firstoftwo
 \else \expandafter \@secondoftwo
 \fi
}%
\providecommand \natexlab [1]{#1}%
\providecommand \enquote  [1]{``#1''}%
\providecommand \bibnamefont  [1]{#1}%
\providecommand \bibfnamefont [1]{#1}%
\providecommand \citenamefont [1]{#1}%
\providecommand \href@noop [0]{\@secondoftwo}%
\providecommand \href [0]{\begingroup \@sanitize@url \@href}%
\providecommand \@href[1]{\@@startlink{#1}\@@href}%
\providecommand \@@href[1]{\endgroup#1\@@endlink}%
\providecommand \@sanitize@url [0]{\catcode `\\12\catcode `\$12\catcode
  `\&12\catcode `\#12\catcode `\^12\catcode `\_12\catcode `\%12\relax}%
\providecommand \@@startlink[1]{}%
\providecommand \@@endlink[0]{}%
\providecommand \url  [0]{\begingroup\@sanitize@url \@url }%
\providecommand \@url [1]{\endgroup\@href {#1}{\urlprefix }}%
\providecommand \urlprefix  [0]{URL }%
\providecommand \Eprint [0]{\href }%
\providecommand \doibase [0]{https://doi.org/}%
\providecommand \selectlanguage [0]{\@gobble}%
\providecommand \bibinfo  [0]{\@secondoftwo}%
\providecommand \bibfield  [0]{\@secondoftwo}%
\providecommand \translation [1]{[#1]}%
\providecommand \BibitemOpen [0]{}%
\providecommand \bibitemStop [0]{}%
\providecommand \bibitemNoStop [0]{.\EOS\space}%
\providecommand \EOS [0]{\spacefactor3000\relax}%
\providecommand \BibitemShut  [1]{\csname bibitem#1\endcsname}%
\let\auto@bib@innerbib\@empty
%</preamble>
\bibitem [{\citenamefont {Alivisatos}(1996)}]{Alivisatos1996}%
  \BibitemOpen
  \bibfield  {author} {\bibinfo {author} {\bibfnamefont {A.~P.}\ \bibnamefont
  {Alivisatos}},\ }\href@noop {} {\bibfield  {journal} {\bibinfo  {journal}
  {Science}\ }\textbf {\bibinfo {volume} {271}},\ \bibinfo {pages} {933}
  (\bibinfo {year} {1996})}
 \BibitemShut {NoStop}%
\bibitem [{\citenamefont {Kubo}(1962)}]{Kubo1962}%
  \BibitemOpen
  \bibfield  {author} {\bibinfo {author} {\bibfnamefont {R.}~\bibnamefont
  {Kubo}},\ }\href@noop {} {\bibfield  {journal} {\bibinfo  {journal} {J. Phys.
  Soc. Jpn.}\ }\textbf {\bibinfo {volume} {17}},\ \bibinfo {pages} {975}
  (\bibinfo {year} {1962})}\BibitemShut {NoStop}%
\bibitem [{\citenamefont {Kubo}\ \emph {et~al.}(1984)\citenamefont {Kubo},
  \citenamefont {Kawabata},\ and\ \citenamefont {Kobayashi}}]{KuboReview1984}%
  \BibitemOpen
  \bibfield  {author} {\bibinfo {author} {\bibfnamefont {R.}~\bibnamefont
  {Kubo}}, \bibinfo {author} {\bibfnamefont {A.}~\bibnamefont {Kawabata}},\
  and\ \bibinfo {author} {\bibfnamefont {S.}~\bibnamefont {Kobayashi}},\
  }\href@noop {} {\bibfield  {journal} {\bibinfo  {journal} {Ann. Rev. Mater.
  Sci.}\ }\textbf {\bibinfo {volume} {14}},\ \bibinfo {pages} {49} (\bibinfo
  {year} {1984})}\BibitemShut {NoStop}%
\bibitem [{\citenamefont {Crespo}\ \emph {et~al.}(2004)\citenamefont {Crespo},
  \citenamefont {Litr{\'a}n}, \citenamefont {Rojas}, \citenamefont {Multigner},
  \citenamefont {De~la Fuente}, \citenamefont {S{\'a}nchez-L{\'o}pez},
  \citenamefont {Garc{\'\i}a}, \citenamefont {Hernando}, \citenamefont
  {Penad{\'e}s},\ and\ \citenamefont {Fern{\'a}ndez}}]{Crespo2004}%
  \BibitemOpen
  \bibfield  {author} {\bibinfo {author} {\bibfnamefont {P.}~\bibnamefont
  {Crespo}}, \bibinfo {author} {\bibfnamefont {R.}~\bibnamefont {Litr{\'a}n}},
  \bibinfo {author} {\bibfnamefont {T.~C.}~\bibnamefont {Rojas}}, \bibinfo
  {author} {\bibfnamefont {M.}~\bibnamefont {Multigner}}, \bibinfo {author}
  {\bibfnamefont {J.~M.}~\bibnamefont {de~la Fuente}}, \bibinfo {author}
  {\bibfnamefont {J.~C.}~\bibnamefont {S{\'a}nchez-L{\'o}pez}}, \bibinfo {author}
  {\bibfnamefont {M.~A.}~\bibnamefont {Garc{\'\i}a}}, \bibinfo {author}
  {\bibfnamefont {A.}~\bibnamefont {Hernando}}, \bibinfo {author}
  {\bibfnamefont {S.}~\bibnamefont {Penad{\'e}s}},\ and\ \bibinfo {author}
  {\bibfnamefont {A.}~\bibnamefont {Fern{\'a}ndez}},\ }\href@noop {} {\bibfield
   {journal} {\bibinfo  {journal} {Phys. Rev. Lett.}\ }\textbf {\bibinfo
  {volume} {93}},\ \bibinfo {pages} {087204} (\bibinfo {year}
  {2004})}\BibitemShut {NoStop}%
\bibitem [{\citenamefont {Garcia}\ \emph {et~al.}(2007)\citenamefont {Garcia},
  \citenamefont {Merino}, \citenamefont {Fern{\'a}ndez~Pinel}, \citenamefont
  {Quesada}, \citenamefont {De~la Venta}, \citenamefont
  {Ru{\'\i}z~Gonz{\'a}lez}, \citenamefont {Castro}, \citenamefont {Crespo},
  \citenamefont {Llopis}, \citenamefont {Gonz{\'a}lez-Calbet} \emph
  {et~al.}}]{Garcia2007}%
  \BibitemOpen
  \bibfield  {author} {\bibinfo {author} {\bibfnamefont {M.}~\bibnamefont
  {Garcia}}, \bibinfo {author} {\bibfnamefont {J.}~\bibnamefont {Merino}},
  \bibinfo {author} {\bibfnamefont {E.}~\bibnamefont {Fern{\'a}ndez~Pinel}},
  \bibinfo {author} {\bibfnamefont {A.}~\bibnamefont {Quesada}}, \bibinfo
  {author} {\bibfnamefont {J.}~\bibnamefont {De~la Venta}}, \bibinfo {author}
  {\bibfnamefont {M.}~\bibnamefont {Ru{\'\i}z~Gonz{\'a}lez}}, \bibinfo {author}
  {\bibfnamefont {G.}~\bibnamefont {Castro}}, \bibinfo {author} {\bibfnamefont
  {P.}~\bibnamefont {Crespo}}, \bibinfo {author} {\bibfnamefont
  {J.}~\bibnamefont {Llopis}}, \bibinfo {author} {\bibfnamefont
  {J.}~\bibnamefont {Gonz{\'a}lez-Calbet}}, \emph {et~al.},\ }\href@noop {}
  {\bibfield  {journal} {\bibinfo  {journal} {Nano Lett.}\ }\textbf {\bibinfo
  {volume} {7}},\ \bibinfo {pages} {1489} (\bibinfo {year} {2007})}\BibitemShut
  {NoStop}%
\bibitem [{\citenamefont {Slichter}(1986)}]{Slicter1986}%
  \BibitemOpen
  \bibfield  {author} {\bibinfo {author} {\bibfnamefont {C.~P.}\ \bibnamefont
  {Slichter}},\ }\href@noop {} {\bibfield  {journal} {\bibinfo  {journal} {Ann.
  Rev. Phys. Chem.}\ }\textbf {\bibinfo {volume} {37}},\ \bibinfo {pages} {25}
  (\bibinfo {year} {1986})}\BibitemShut {NoStop}%
\bibitem [{\citenamefont {Halperin}(1986)}]{Halperin1986}%
  \BibitemOpen
  \bibfield  {author} {\bibinfo {author} {\bibfnamefont {W.~P.}\ \bibnamefont
  {Halperin}},\ }\href {https://doi.org/10.1103/RevModPhys.58.533} {\bibfield
  {journal} {\bibinfo  {journal} {Rev. Mod. Phys.}\ }\textbf {\bibinfo {volume}
  {58}},\ \bibinfo {pages} {533} (\bibinfo {year} {1986})}\BibitemShut
  {NoStop}%
\bibitem [{\citenamefont {Goto}\ \emph {et~al.}(1989)\citenamefont {Goto},
  \citenamefont {Komori},\ and\ \citenamefont {Kobayashi}}]{Goto1989}%
  \BibitemOpen
  \bibfield  {author} {\bibinfo {author} {\bibfnamefont {T.}~\bibnamefont
  {Goto}}, \bibinfo {author} {\bibfnamefont {F.}~\bibnamefont {Komori}},\ and\
  \bibinfo {author} {\bibfnamefont {S.-i.}\ \bibnamefont {Kobayashi}},\ }\href
  {https://doi.org/10.1143/JPSJ.58.3788} {\bibfield  {journal} {\bibinfo
  {journal} {J. Phys. Soc. Jpn.}\ }\textbf {\bibinfo {volume} {58}},\ \bibinfo
  {pages} {3788} (\bibinfo {year} {1989})}\BibitemShut {NoStop}%
\bibitem [{\citenamefont {van~der Klink}\ and\ \citenamefont
  {Brom}(2000)}]{Klink2000}%
  \BibitemOpen
  \bibfield  {author} {\bibinfo {author} {\bibfnamefont {J.~J.}\ \bibnamefont
  {van~der Klink}}\ and\ \bibinfo {author} {\bibfnamefont {H.~B.}\ \bibnamefont
  {Brom}},\ }\href@noop {} {\bibfield  {journal} {\bibinfo  {journal} {Prog.
  Nucl. Magn. Reson. Spectrosc.}\ }\textbf {\bibinfo {volume} {36}},\ \bibinfo
  {pages} {89} (\bibinfo {year} {2000})}\BibitemShut {NoStop}%
\bibitem [{\citenamefont {Marbella}\ and\ \citenamefont
  {Millstone}(2015)}]{Marbella2015}%
  \BibitemOpen
  \bibfield  {author} {\bibinfo {author} {\bibfnamefont {L.~E.}\ \bibnamefont
  {Marbella}}\ and\ \bibinfo {author} {\bibfnamefont {J.~E.}\ \bibnamefont
  {Millstone}},\ }\href@noop {} {\bibfield  {journal} {\bibinfo  {journal}
  {Chem. Mater.}\ }\textbf {\bibinfo {volume} {27}},\ \bibinfo {pages} {2721}
  (\bibinfo {year} {2015})}\BibitemShut {NoStop}%
\bibitem [{\citenamefont {Seh}\ \emph {et~al.}(2017)\citenamefont {Seh},
  \citenamefont {Kibsgaard}, \citenamefont {Dickens}, \citenamefont
  {Chorkendorff}, \citenamefont {N{\o}rskov},\ and\ \citenamefont
  {Jaramillo}}]{Seh2017}%
  \BibitemOpen
  \bibfield  {author} {\bibinfo {author} {\bibfnamefont {Z.~W.}\ \bibnamefont
  {Seh}}, \bibinfo {author} {\bibfnamefont {J.}~\bibnamefont {Kibsgaard}},
  \bibinfo {author} {\bibfnamefont {C.~F.}\ \bibnamefont {Dickens}}, \bibinfo
  {author} {\bibfnamefont {I.}~\bibnamefont {Chorkendorff}}, \bibinfo {author}
  {\bibfnamefont {J.~K.}\ \bibnamefont {N{\o}rskov}},\ and\ \bibinfo {author}
  {\bibfnamefont {T.~F.}\ \bibnamefont {Jaramillo}},\ }\href@noop {} {\bibfield
   {journal} {\bibinfo  {journal} {Science}\ }\textbf {\bibinfo {volume}
  {355}},\ \bibinfo {pages} {eaad4998} (\bibinfo {year} {2017})}\BibitemShut
  {NoStop}%
\bibitem [{\citenamefont {Slichter}(1981)}]{Slichter1981}%
  \BibitemOpen
  \bibfield  {author} {\bibinfo {author} {\bibfnamefont {C.~P.}\ \bibnamefont
  {Slichter}},\ }\href@noop {} {\bibfield  {journal} {\bibinfo  {journal}
  {Surf. Sci.}\ }\textbf {\bibinfo {volume} {106}},\ \bibinfo {pages} {382}
  (\bibinfo {year} {1981})}\BibitemShut {NoStop}%
\bibitem [{\citenamefont {Rhodes}\ \emph
  {et~al.}(1982{\natexlab{a}})\citenamefont {Rhodes}, \citenamefont {Wang},
  \citenamefont {Stokes}, \citenamefont {Slichter},\ and\ \citenamefont
  {Sinfelt}}]{Rhodes1982i}%
  \BibitemOpen
  \bibfield  {author} {\bibinfo {author} {\bibfnamefont {H.~E.}\ \bibnamefont
  {Rhodes}}, \bibinfo {author} {\bibfnamefont {P.-K.}\ \bibnamefont {Wang}},
  \bibinfo {author} {\bibfnamefont {H.~T.}\ \bibnamefont {Stokes}}, \bibinfo
  {author} {\bibfnamefont {C.~P.}\ \bibnamefont {Slichter}},\ and\ \bibinfo
  {author} {\bibfnamefont {J.~H.}~\bibnamefont {Sinfelt}},\ }\href@noop {}
  {\bibfield  {journal} {\bibinfo  {journal} {Phys. Rev. B}\ }\textbf {\bibinfo
  {volume} {26}},\ \bibinfo {pages} {3559} (\bibinfo {year}
  {1982}{\natexlab{a}})}\BibitemShut {NoStop}%
\bibitem [{\citenamefont {Rhodes}\ \emph
  {et~al.}(1982{\natexlab{b}})\citenamefont {Rhodes}, \citenamefont {Wang},
  \citenamefont {Makowka}, \citenamefont {Rudaz}, \citenamefont {Stokes},
  \citenamefont {Slichter},\ and\ \citenamefont {Sinfelt}}]{Rhodes1982ii}%
  \BibitemOpen
  \bibfield  {author} {\bibinfo {author} {\bibfnamefont {H.~E.}\ \bibnamefont
  {Rhodes}}, \bibinfo {author} {\bibfnamefont {P.-K.}\ \bibnamefont {Wang}},
  \bibinfo {author} {\bibfnamefont {C.~D.}\ \bibnamefont {Makowka}}, \bibinfo
  {author} {\bibfnamefont {S.~L.}\ \bibnamefont {Rudaz}}, \bibinfo {author}
  {\bibfnamefont {H.~T.}\ \bibnamefont {Stokes}}, \bibinfo {author}
  {\bibfnamefont {C.~P.}\ \bibnamefont {Slichter}},\ and\ \bibinfo {author}
  {\bibfnamefont {J.~H.}~\bibnamefont {Sinfelt}},\ }\href@noop {} {\bibfield
  {journal} {\bibinfo  {journal} {Phys. Rev. B}\ }\textbf {\bibinfo {volume}
  {26}},\ \bibinfo {pages} {3569} (\bibinfo {year}
  {1982}{\natexlab{b}})}\BibitemShut {NoStop}%
\bibitem [{\citenamefont {Makowka}\ \emph {et~al.}(1982)\citenamefont
  {Makowka}, \citenamefont {Slichter},\ and\ \citenamefont
  {Sinfelt}}]{Makowka1982}%
  \BibitemOpen
  \bibfield  {author} {\bibinfo {author} {\bibfnamefont {C.~D.}\ \bibnamefont
  {Makowka}}, \bibinfo {author} {\bibfnamefont {C.~P.}\ \bibnamefont
  {Slichter}},\ and\ \bibinfo {author} {\bibfnamefont {J.~H.}~\bibnamefont
  {Sinfelt}},\ }\href@noop {} {\bibfield  {journal} {\bibinfo  {journal} {Phys.
  Rev. Lett.}\ }\textbf {\bibinfo {volume} {49}},\ \bibinfo {pages} {379}
  (\bibinfo {year} {1982})}\BibitemShut {NoStop}%
\bibitem [{\citenamefont {Van~der Klink}\ \emph {et~al.}(1984)\citenamefont
  {Van~der Klink}, \citenamefont {Buttet},\ and\ \citenamefont
  {Graetzel}}]{Klink1984}%
  \BibitemOpen
  \bibfield  {author} {\bibinfo {author} {\bibfnamefont {J.~J.}~\bibnamefont
  {van~der Klink}}, \bibinfo {author} {\bibfnamefont {J.}~\bibnamefont
  {Buttet}},\ and\ \bibinfo {author} {\bibfnamefont {M.}~\bibnamefont
  {Graetzel}},\ }\href@noop {} {\bibfield  {journal} {\bibinfo  {journal}
  {Phys. Rev. B}\ }\textbf {\bibinfo {volume} {29}},\ \bibinfo {pages} {6352}
  (\bibinfo {year} {1984})}\BibitemShut {NoStop}%
\bibitem [{\citenamefont {Makowka}\ \emph {et~al.}(1985)\citenamefont
  {Makowka}, \citenamefont {Slichter},\ and\ \citenamefont
  {Sinfelt}}]{Makowka1985}%
  \BibitemOpen
  \bibfield  {author} {\bibinfo {author} {\bibfnamefont {C.~D.}\ \bibnamefont
  {Makowka}}, \bibinfo {author} {\bibfnamefont {C.~P.}\ \bibnamefont
  {Slichter}},\ and\ \bibinfo {author} {\bibfnamefont {J.~H.}~\bibnamefont
  {Sinfelt}},\ }\href@noop {} {\bibfield  {journal} {\bibinfo  {journal} {Phys.
  Rev. B}\ }\textbf {\bibinfo {volume} {31}},\ \bibinfo {pages} {5663}
  (\bibinfo {year} {1985})}\BibitemShut {NoStop}%
\bibitem [{\citenamefont {Tong}\ \emph {et~al.}(1993)\citenamefont {Tong},
  \citenamefont {Van~der Klink}, \citenamefont {Clugnet}, \citenamefont
  {Renouprez}, \citenamefont {Laub},\ and\ \citenamefont {Buffat}}]{Tong1993}%
  \BibitemOpen
  \bibfield  {author} {\bibinfo {author} {\bibfnamefont {Y.~Y.}~\bibnamefont
  {Tong}}, \bibinfo {author} {\bibfnamefont {J.~J.}~\bibnamefont {van~der Klink}},
  \bibinfo {author} {\bibfnamefont {G.}~\bibnamefont {Clugnet}}, \bibinfo
  {author} {\bibfnamefont {A.}~\bibnamefont {Renouprez}}, \bibinfo {author}
  {\bibfnamefont {D.}~\bibnamefont {Laub}},\ and\ \bibinfo {author}
  {\bibfnamefont {P.}~\bibnamefont {Buffat}},\ }\href@noop {} {\bibfield
  {journal} {\bibinfo  {journal} {Surf. Sci.}\ }\textbf {\bibinfo {volume}
  {292}},\ \bibinfo {pages} {276} (\bibinfo {year} {1993})}\BibitemShut
  {NoStop}%
\bibitem [{\citenamefont {Tong}\ \emph {et~al.}(1995)\citenamefont {Tong},
  \citenamefont {Laub}, \citenamefont {Schulz-Ekloff}, \citenamefont
  {Renouprez},\ and\ \citenamefont {Van~der Klink}}]{Tong1995}%
  \BibitemOpen
  \bibfield  {author} {\bibinfo {author} {\bibfnamefont {Y.~Y.}~\bibnamefont
  {Tong}}, \bibinfo {author} {\bibfnamefont {D.}~\bibnamefont {Laub}}, \bibinfo
  {author} {\bibfnamefont {G.}~\bibnamefont {Schulz-Ekloff}}, \bibinfo {author}
  {\bibfnamefont {A.~J.}~\bibnamefont {Renouprez}},\ and\ \bibinfo {author}
  {\bibfnamefont {J.~J.}~\bibnamefont {van~der Klink}},\ }\href@noop {} {\bibfield
   {journal} {\bibinfo  {journal} {Phys. Rev. B}\ }\textbf {\bibinfo {volume}
  {52}},\ \bibinfo {pages} {8407} (\bibinfo {year} {1995})}\BibitemShut
  {NoStop}%
\bibitem [{\citenamefont {Van Der~Putten}\ \emph {et~al.}(1993)\citenamefont
  {Van Der~Putten}, \citenamefont {Brom}, \citenamefont {Witteveen},
  \citenamefont {De~Jongh},\ and\ \citenamefont {Schmid}}]{Putten1993}%
  \BibitemOpen
  \bibfield  {author} {\bibinfo {author} {\bibfnamefont {D.}~\bibnamefont {Van
  Der~Putten}}, \bibinfo {author} {\bibfnamefont {H.}~\bibnamefont {Brom}},
  \bibinfo {author} {\bibfnamefont {J.}~\bibnamefont {Witteveen}}, \bibinfo
  {author} {\bibfnamefont {L.}~\bibnamefont {De~Jongh}},\ and\ \bibinfo
  {author} {\bibfnamefont {G.}~\bibnamefont {Schmid}},\ }\href@noop {}
  {\bibfield  {journal} {\bibinfo  {journal} {Z. Phys. D}\ }\textbf {\bibinfo
  {volume} {26}},\ \bibinfo {pages} {21} (\bibinfo {year} {1993})}\BibitemShut
  {NoStop}%
  
  \bibitem [{\citenamefont {Yu}\ and\ \citenamefont {Yu}(2004)}]{Yu1980}%
  \BibitemOpen
  \bibfield  {author} {\bibinfo {author} {\bibfnamefont {I.}~\bibnamefont
  {Yu}}\ ,\ \bibinfo {author} {\bibfnamefont {AAV.}~\bibnamefont {Gibson}},
  \bibinfo {author} {\bibfnamefont {E.R.}~\bibnamefont {Hunt}}\ and\
  \bibinfo {author} {\bibfnamefont {W.P.}~\bibnamefont {Halperin}},
  }\href@noop {} {\bibfield  {journal} {\bibinfo  {journal} {Phys. Rev. Lett}\
  }\textbf {\bibinfo {volume} {44}},\ \bibinfo {pages} {348} (\bibinfo {year}
  {1980})}\BibitemShut {NoStop}%
  
    \bibitem [{\citenamefont {Yu}\ and\ \citenamefont {Yu}(1993)}]{Yu1993}%
  \BibitemOpen
  \bibfield  {author} {\bibinfo {author} {\bibfnamefont {I.}~\bibnamefont
  {Yu}}\ ,\ \bibinfo {author} {\bibfnamefont {W.P.}~\bibnamefont {Halperin}},
  }\href@noop {} {\bibfield  {journal} {\bibinfo  {journal} {Phys. Rev. B}\
  }\textbf {\bibinfo {volume} {47}},\ \bibinfo {pages} {15830} (\bibinfo {year}
  {1993})}\BibitemShut {NoStop}%


  
  
\bibitem [{\citenamefont {Fritschij}\ \emph {et~al.}(1999)\citenamefont
  {Fritschij}, \citenamefont {Brom}, \citenamefont {de~Jongh},\ and\
  \citenamefont {Schmid}}]{Fritschij1999}%
  \BibitemOpen
  \bibfield  {author} {\bibinfo {author} {\bibfnamefont {F.~C.}\ \bibnamefont
  {Fritschij}}, \bibinfo {author} {\bibfnamefont {H.~B.}\ \bibnamefont {Brom}},
  \bibinfo {author} {\bibfnamefont {L.~J.}\ \bibnamefont {de~Jongh}},\ and\
  \bibinfo {author} {\bibfnamefont {G.}~\bibnamefont {Schmid}},\ }\href
  {https://doi.org/10.1103/PhysRevLett.82.2167} {\bibfield  {journal} {\bibinfo
   {journal} {Phys. Rev. Lett.}\ }\textbf {\bibinfo {volume} {82}},\ \bibinfo
  {pages} {2167} (\bibinfo {year} {1999})}\BibitemShut {NoStop}%
\bibitem [{\citenamefont {Rees}\ \emph {et~al.}(2013)\citenamefont {Rees},
  \citenamefont {Orr}, \citenamefont {Barrett}, \citenamefont {Fisher},
  \citenamefont {Houghton}, \citenamefont {Spikes}, \citenamefont {Theobald},
  \citenamefont {Thompsett}, \citenamefont {Smith},\ and\ \citenamefont
  {Hanna}}]{Rees2013}%
  \BibitemOpen
  \bibfield  {author} {\bibinfo {author} {\bibfnamefont {G.~J.}\ \bibnamefont
  {Rees}}, \bibinfo {author} {\bibfnamefont {S.~T.}\ \bibnamefont {Orr}},
  \bibinfo {author} {\bibfnamefont {L.~O.}\ \bibnamefont {Barrett}}, \bibinfo
  {author} {\bibfnamefont {J.~M.}\ \bibnamefont {Fisher}}, \bibinfo {author}
  {\bibfnamefont {J.}~\bibnamefont {Houghton}}, \bibinfo {author}
  {\bibfnamefont {G.~H.}\ \bibnamefont {Spikes}}, \bibinfo {author}
  {\bibfnamefont {B.~R.}\ \bibnamefont {Theobald}}, \bibinfo {author}
  {\bibfnamefont {D.}~\bibnamefont {Thompsett}}, \bibinfo {author}
  {\bibfnamefont {M.~E.}\ \bibnamefont {Smith}},\ and\ \bibinfo {author}
  {\bibfnamefont {J.~V.}\ \bibnamefont {Hanna}},\ }\href@noop {} {\bibfield
  {journal} {\bibinfo  {journal} {Phys. Chem. Chem. Phys.}\ }\textbf {\bibinfo
  {volume} {15}},\ \bibinfo {pages} {17195} (\bibinfo {year}
  {2013})}\BibitemShut {NoStop}%
\bibitem [{\citenamefont {Okuno}\ \emph {et~al.}(2020)\citenamefont {Okuno},
  \citenamefont {Manago}, \citenamefont {Kitagawa}, \citenamefont {Ishida},
  \citenamefont {Kusada},\ and\ \citenamefont {Kitagawa}}]{Okuno2020}%
  \BibitemOpen
  \bibfield  {author} {\bibinfo {author} {\bibfnamefont {T.}~\bibnamefont
  {Okuno}}, \bibinfo {author} {\bibfnamefont {M.}~\bibnamefont {Manago}},
  \bibinfo {author} {\bibfnamefont {S.}~\bibnamefont {Kitagawa}}, \bibinfo
  {author} {\bibfnamefont {K.}~\bibnamefont {Ishida}}, \bibinfo {author}
  {\bibfnamefont {K.}~\bibnamefont {Kusada}},\ and\ \bibinfo {author}
  {\bibfnamefont {H.}~\bibnamefont {Kitagawa}},\ }\href
  {https://doi.org/10.1103/PhysRevB.101.121406} {\bibfield  {journal} {\bibinfo
   {journal} {Phys. Rev. B}\ }\textbf {\bibinfo {volume} {101}},\ \bibinfo
  {pages} {121406(R)} (\bibinfo {year} {2020})}\BibitemShut {NoStop}%
\bibitem [{\citenamefont {Chen}\ and\ \citenamefont
  {Kimura}(2001)}]{ChenKimura2001}%
  \BibitemOpen
  \bibfield  {author} {\bibinfo {author} {\bibfnamefont {S.}~\bibnamefont
  {Chen}}\ and\ \bibinfo {author} {\bibfnamefont {K.}~\bibnamefont {Kimura}},\
  }\href {https://doi.org/10.1021/jp0037798} {\bibfield  {journal} {\bibinfo
  {journal} {J. Phys. Chem. B}\ }\textbf {\bibinfo {volume} {105}},\ \bibinfo
  {pages} {5397} (\bibinfo {year} {2001})}\BibitemShut {NoStop}%
\bibitem [{\citenamefont {Tong}\ \emph {et~al.}(2005)\citenamefont {Tong},
  \citenamefont {Zelakiewicz}, \citenamefont {Dy},\ and\ \citenamefont
  {Pogozelski}}]{Tong2005}%
  \BibitemOpen
  \bibfield  {author} {\bibinfo {author} {\bibfnamefont {Y.}~\bibnamefont
  {Tong}}, \bibinfo {author} {\bibfnamefont {B.~S.}\ \bibnamefont
  {Zelakiewicz}}, \bibinfo {author} {\bibfnamefont {B.~M.}\ \bibnamefont
  {Dy}},\ and\ \bibinfo {author} {\bibfnamefont {A.~R.}\ \bibnamefont
  {Pogozelski}},\ }\href@noop {} {\bibfield  {journal} {\bibinfo  {journal}
  {Chem. Phys. Lett.}\ }\textbf {\bibinfo {volume} {406}},\ \bibinfo {pages}
  {137} (\bibinfo {year} {2005})}\BibitemShut {NoStop}%
\bibitem [{\citenamefont {Bucher}\ \emph {et~al.}(1989)\citenamefont {Bucher},
  \citenamefont {Buttet}, \citenamefont {Van~der Klink},\ and\ \citenamefont
  {Graetzel}}]{Bucher1989}%
  \BibitemOpen
  \bibfield  {author} {\bibinfo {author} {\bibfnamefont {J.}~\bibnamefont
  {Bucher}}, \bibinfo {author} {\bibfnamefont {J.}~\bibnamefont {Buttet}},
  \bibinfo {author} {\bibfnamefont {J.~J.}~\bibnamefont {van~der Klink}},\ and\
  \bibinfo {author} {\bibfnamefont {M.}~\bibnamefont {Graetzel}},\ }\href@noop
  {} {\bibfield  {journal} {\bibinfo  {journal} {Surf. Sci.}\ }\textbf
  {\bibinfo {volume} {214}},\ \bibinfo {pages} {347} (\bibinfo {year}
  {1989})}\BibitemShut {NoStop}%
\bibitem [{\citenamefont {Carter}\ \emph {et~al.}(1977)\citenamefont {Carter},
  \citenamefont {Bennett},\ and\ \citenamefont {Kahan}}]{MetallicShifts}%
  \BibitemOpen
  \bibfield  {author} {\bibinfo {author} {\bibfnamefont {G.}~\bibnamefont
  {Carter}}, \bibinfo {author} {\bibfnamefont {L.}~\bibnamefont {Bennett}},\
  and\ \bibinfo {author} {\bibfnamefont {D.}~\bibnamefont {Kahan}},\ }\href
  {https://books.google.co.jp/books?id=NylPAQAAIAAJ} {}Progress in materials
  science\ (\bibinfo  {publisher} {Pergamon Press},\ \bibinfo {year}
  {1977})\BibitemShut {NoStop}%
\bibitem [{\citenamefont {Shaham}\ \emph {et~al.}(1978)\citenamefont {Shaham},
  \citenamefont {El-Hanany},\ and\ \citenamefont {Zamir}}]{Shaham1978}%
  \BibitemOpen
  \bibfield  {author} {\bibinfo {author} {\bibfnamefont {M.}~\bibnamefont
  {Shaham}}, \bibinfo {author} {\bibfnamefont {U.}~\bibnamefont {El-Hanany}},\
  and\ \bibinfo {author} {\bibfnamefont {D.}~\bibnamefont {Zamir}},\
  }\href@noop {} {\bibfield  {journal} {\bibinfo  {journal} {Phys. Rev. B}\
  }\textbf {\bibinfo {volume} {17}},\ \bibinfo {pages} {3513} (\bibinfo {year}
  {1978})}\BibitemShut {NoStop}%
\bibitem [{\citenamefont {Bucher}\ and\ \citenamefont {Van~der
  Klink}(1988)}]{Bucher1988}%
  \BibitemOpen
  \bibfield  {author} {\bibinfo {author} {\bibfnamefont {J.~P.}~\bibnamefont
  {Bucher}}\ and\ \bibinfo {author} {\bibfnamefont {J.~J.}~\bibnamefont {van~der
  Klink}},\ }\href@noop {} {\bibfield  {journal} {\bibinfo  {journal} {Phys.
  Rev. B}\ }\textbf {\bibinfo {volume} {38}},\ \bibinfo {pages} {11038}
  (\bibinfo {year} {1988})}\BibitemShut {NoStop}%

\bibitem [{\citenamefont {Bloembergen}(1954)}]{Bloembergen1954}%
  \BibitemOpen
  \bibfield  {author} {\bibinfo {author} {\bibfnamefont {N.}~\bibnamefont
  {Bloembergen}},\ }\href@noop {} {\bibfield  {journal} {\bibinfo  {journal}
  {Physica}\ }\textbf {\bibinfo {volume} {20}},\ \bibinfo {pages} {1130}
  (\bibinfo {year} {1954})}\BibitemShut {NoStop}%


\bibitem [{\citenamefont {Selbach}(1979)}]{Selbach1979}%
  \BibitemOpen
    \bibfield  {author} {\bibinfo {author} {\bibfnamefont {H.}~\bibnamefont
  {Selbach}}, \bibinfo {author} {\bibfnamefont {O.}~\bibnamefont {Kanert}},
  \bibinfo {author} {\bibfnamefont {D.}~\bibnamefont {Wolf}},\ } 
  \href@noop {}{\bibfield  {journal} {\bibinfo {journal}  {Phys. Rev. B}
  }\textbf {\bibinfo {volume} {19}},\ \bibinfo {pages} {4435}
  (\bibinfo {year} {1979})}\BibitemShut {NoStop}%





\bibitem [{\citenamefont {Sigalas}\ \emph {et~al.}(1992)\citenamefont
  {Sigalas}, \citenamefont {Papaconstantopoulos},\ and\ \citenamefont
  {Bacalis}}]{Sigalas1992}%
  \BibitemOpen
  \bibfield  {author} {\bibinfo {author} {\bibfnamefont {M.}~\bibnamefont
  {Sigalas}}, \bibinfo {author} {\bibfnamefont {D.~A.}\ \bibnamefont
  {Papaconstantopoulos}},\ and\ \bibinfo {author} {\bibfnamefont {N.~C.}\
  \bibnamefont {Bacalis}},\ }\href {https://doi.org/10.1103/PhysRevB.45.5777}
  {\bibfield  {journal} {\bibinfo  {journal} {Phys. Rev. B}\ }\textbf {\bibinfo
  {volume} {45}},\ \bibinfo {pages} {5777} (\bibinfo {year}
  {1992})}\BibitemShut {NoStop}%
\bibitem [{\citenamefont {Xiao}\ and\ \citenamefont {Wang}(2004)}]{Xiao2004}%
  \BibitemOpen
  \bibfield  {author} {\bibinfo {author} {\bibfnamefont {L.}~\bibnamefont
  {Xiao}}\ and\ \bibinfo {author} {\bibfnamefont {L.}~\bibnamefont {Wang}},\
  }\href@noop {} {\bibfield  {journal} {\bibinfo  {journal} {J. Phys. Chem. A}\
  }\textbf {\bibinfo {volume} {108}},\ \bibinfo {pages} {8605} (\bibinfo {year}
  {2004})}\BibitemShut {NoStop}%
  
  \bibitem [{\citenamefont {Fresch}(2014)}]{Fresch}%
    \BibitemOpen
  \bibfield  {author} {\bibinfo {author} {\bibfnamefont {B.}\ \bibnamefont
  {Fresch}}, \bibinfo {author} {\bibfnamefont {F.}\ \bibnamefont
  {Remacle}},\ }\href@noop {} {\bibfield  {journal} {\bibinfo  {journal} {J. Phys. Chem. C}\ }\textbf {\bibinfo {volume} {118}},\ \bibinfo {pages} {9790}
  (\bibinfo {year} {2014})}\BibitemShut {NoStop}%

  
  
    
  
\end{thebibliography}
\end{document}